\documentstyle[12pt]{article}
\makeatletter
\def\fmslash{\@ifnextchar[{\fmsl@sh}{\fmsl@sh[0mu]}}
\def\fmsl@sh[#1]#2{%
  \mathchoice
    {\@fmsl@sh\displaystyle{#1}{#2}}%
    {\@fmsl@sh\textstyle{#1}{#2}}%
    {\@fmsl@sh\scriptstyle{#1}{#2}}%
    {\@fmsl@sh\scriptscriptstyle{#1}{#2}}}
\def\@fmsl@sh#1#2#3{\m@th\ooalign{$\hfil#1\mkern#2/\hfil$\crcr$#1#3$}}
\makeatother
\begin{document}
\thispagestyle{empty}
\begin{titlepage}
\hskip -3cm
\begin{flushright}
{\bf TTP 96 -- 56}
\end{flushright}


\begin{center}
{\Large\bf Review of Heavy Quark Effective Theory}
\end{center}

\vspace{0.8cm}

\begin{center}
{\sc Thomas Mannel}  \vspace*{2mm} \\
{\sl Institut f\"{u}r Theoretische Teilchenphysik,
     University of Karlsruhe \vspace*{1mm} \\
     D -- 76128 Karlsruhe, Germany.}
\vspace*{2cm} \\
{\it Contribution to the Workshop on HEAVY QUAKRS AT FIXED TARGET,  \\
     Rheinfals Castle, St.\ Goar, Germany, October 3-6, 1996}
\end{center}
\vfill
\begin{abstract}
\noindent
A short review of a few selected topics in Heavy Quark Effective Theory 
is given. Applications to exclusive decays are discussed. 
\end{abstract}
\end{titlepage}
\newpage
\section{Introduction}
Heavy Quark Effective Theory (HQET) or, more generally, the expansion 
in inverse powers of the heavy quark mass $m_Q$, has become a generally 
accepted and widely used tool in heavy quark physics\cite{HQET}. 
Based on the infinite
mass limit $m_Q \to \infty$ of QCD it provides a model independent starting
point for the description of weak transitions involving heavy quarks.   

The idea to exploit the fact that the mass of a heavy quark is large 
compared to the typical scale $\Lambda$ of the light QCD degrees of freedom
(e.g.\ the constituent mass of a light quark or the scale of the 
QCD coupling constant $\Lambda_{QCD}$) is in fact quite old\cite{HQETold}.  
However, in the late eighties a breakthrough was achieved by mainly two
observations. First, in the infinite mass limit QCD exhibits an  
additional flavour symmetry and a spin symmetry\cite{IsgurWise}, 
the group theory of 
which allow model independent statements concerning weak decays of 
heavy hadrons.
Second, it was noted that the $1/m_Q$ expansion of QCD 
can be formulated as an effective field theory \cite{HQEFFT}, 
which allows to access
the corrections to the infinite mass limit in a systematic way. 

Since then the field of heavy quark physics has attracted a lot of 
attention, documented by an enormous number of papers that have been 
published using these methods. In addition, from the experimental side
a large effort is made to investigate the decays of bottom hadrons 
in order to pin down the origin of quark mixing and CP violation. 
In the standard model (SM) all of this is encoded in the CKM matrix, 
and the measurement of those elements of this matrix, which are only 
poorly known, involves mainly weak processes of $b$ quarks.  
   
HQET, or more generally the systematic application of the $1/m_Q$ 
expansion in QCD, has brought some progress in the 
determination of these CKM matrix elements, since heavy quark symmetries
allow a drastic reduction of the hadronic uncertainties which enter
the game through our ignorance to deal with the QCD bound state problem
from first principles. In particular, the $b \to c$ semileptonic decay 
may considered as a heavy $\to$ heavy transition, where heavy quark 
symmetries work very efficiently; consequently, the $1/m_Q$ expansion 
allows for an almost model independent determination of 
the CKM matrix element $V_{cb}$. 

In the next section a brief summary on HQET and heavy quark symmetries 
is given. Section 3 deals with the ``picture book application'', namely  
the exclusive $b \to c $ semileptonic decays. Heavy quark symmetries are
also of some use in the case of a heavy hadron decaying into something 
light, and in section 4 this is considered for the case of heavy to 
light transitions. Finally a summary and a few conclusions are given.

\section{Synopsis of HQET and heavy quark symmetries}
\subsection{Lagrangian and Fields}
HQET is an effective field theory which may be obtained from QCD 
by performing a $1/m_Q$ expansion. The leading term corresponds to 
the infinite mass limit in which the heavy quark acts as a static 
color source. The momentum $p_Q$ of the heavy quark scales with its mass
and in order to perform the infinite mass limit it is convenient 
to use the velocity $v$ of the heavy quark as the basic kinematic 
quantity. To this end the heavy quark momentum is split into a large part 
$m_Q v$ and a residual part $k$, which is assumed not to scale with 
the heavy mass. Thus 
\begin{equation}
p_Q = m_Q v + k = m_Q \left(v + \frac{k}{m_Q} \right)
\end{equation}
We shall consider exclusively hadrons containing only a single heavy quark
such that in the infinite mass limit the velocity $v$ of the heavy quark 
becomes simply the velocity of the heavy hadron. 

In order to write down a field theory which describes the static heavy 
quark one may go through the usual steps of the construction of an effective 
field theory, namely to integrate out the heavy degrees of freedom
\cite{derivation}. An alternative method \cite{koerner}
is to perform a Foldy 
Wouthuysen Transformation as it is used in the standard non-relativistic 
reduction of the Dirac equation. Although the $1/m_Q$ expansion of the 
Lagrangian and the corresponding expansion of the fields look completely 
different, the results for physical matrix elements will be the same
since the $1/m_Q$ expansion of the QCD Greens functions constructed from 
this effective theory has to be unique. 

In the notation of \cite{derivation} one obtains 
\begin{eqnarray} \label{HarvField}
Q(x) &=& e^{-im_Qvx} \left[ 1 + 
         \left( \frac{1}{2m + ivD} \right) i \fmslash{D}_\perp \right] h_v \\ 
     &=& e^{-im_Qvx} \left[ 1 + \frac{1}{2m_Q} \fmslash{D}_\perp +  \nonumber 
\left(\frac{1}{2m_Q}\right)^2 (-ivD) \fmslash{D}_\perp + \cdots \right] h_v \\ 
\label{HarvLag}
{\cal L} &=& \bar{h}_v (iv D) h_v + 
\bar{h}_v i \fmslash{D}_\perp \left( \frac{1}{2m + ivD} \right) 
          i \fmslash{D}_\perp h_v \\
         &=& \bar{h}_v (iv D) h_v 
             + \frac{1}{2m} \bar{h}_v (i\fmslash{D}_\perp)^2 i h_v
          +  \left( \frac{1}{2m}\right)^2 
             \bar{h}_v (i\fmslash{D}_\perp) (-ivD) 
         (i\fmslash{D}_\perp)  h_v + \cdots \nonumber
\end{eqnarray}
where $D$ is the covariant derivative of QCD and 
$Q(x)$ is the heavy quark field in full QCD, while $h_v$ is the 
static heavy quark moving with the velocity $v$. Note that $h_v$ 
corresponds to the upper components of the full field since 
\begin{equation}
P_+ h_v = h_v , \quad  P_- h_v = 0 , \quad 
P_\pm = \frac{1}{2} (\fmslash{v} \pm 1)
\end{equation}

The leading terms of these expansions define the static limit; the 
static lagrangian
\begin{equation} \label{Lstat} 
{\cal L}_{stat} = \bar{h}_v (iv D) h_v 
\end{equation}
is a dimension-four operator and defines (in combination with the 
usual lagrangian for the light degrees of freedom) a renormalizable 
field theory.

\subsection{Heavy Quark Symmetries}

In the case in which the bottom and the charam quark are assumed to be 
heavy one would write a static lagrangian for both quarks  
\begin{equation} \label{Lbc}
{\cal L}_{stat} = \bar{b}_v (v \cdot D) b_v + \bar{c}_{v'} (v \cdot D) c_{v'} ,
\end{equation}
where $b_v$ ($c_{v'}$) is the field operator for the $b$ ($c$) quark 
moving with velocity $v$ ($v'$). In particular, the masses of the heavy 
quarks do not appear in the Lagrangian (\ref{Lbc}), and as a consequence 
(\ref{Lbc}) in the case $v = v'$ exhibits an $SU(2)$ Heavy Flavour 
Symmetry which rotates the $b_v$ field into the $c_v$ field. 

The static heavy quark field $h_v$ is still a two component object 
corresponding to the upper component of the full heavy quark field $Q$. 
However, both spin directions couple in the same way to the gluons; 
we may rewrite the leading-order Lagrangian as 
\begin{equation}
{\cal L} = \bar{h}_v^{+s} (iv D) h_v^{+s} + \bar{h}_v^{-s} (iv D) h_v^{-s},
\end{equation}
where $h_v^{\pm s}$ are the projections of the heavy quark field on a 
definite spin direction $s$
\begin{equation}
h_v^{\pm s} = \frac{1}{2} (1 \pm \gamma_5 \fmslash{s}) h_v, 
\quad s\cdot v = 0 .
\end{equation}
This Lagrangian has a symmetry under the rotations of the heavy quark 
spin and hence all the heavy hadron states moving with the velocity $v$
fall into spin-symmetry doublets as $m_Q \to \infty$. 
The simplest spin-symmetry doublet in the mesonic case consists of the 
pseudoscalar meson $H(v)$ and the corresponding vector meson 
$H^* (v,\epsilon)$, since a $90^\circ$-spin rotation $R(\epsilon)$ around 
the rotation axis $\epsilon$ ($v \epsilon = 0$) yields
\begin{equation}
R(\epsilon) | H (v) \rangle = 
(-i) | H^* (v,\epsilon) \rangle ,
\end{equation}

In the heavy-mass limit the spin symmetry partners have to be 
degenerate and their splitting has to scale as $1/m_Q$. In other 
words, the quantity 
\begin{equation}
\lambda_2 = \frac{1}{4} (M_{H^*}^2 - M_H^2) 
\end{equation}
has to be the same for all spin symmetry doublets of heavy ground state 
mesons. This is well supported by data: For both the $(B,B^*)$ and the 
$(D,D^*)$ doublets one finds a value of $\lambda_2 \sim 0.12$ GeV${}^2$.
This shows that 
the spin-symmetry partners become degenerate in the infinite mass limit and 
the splitting between them scales as $1/m_Q$. 

In the infinite mass limit the symmetries imply relations between matrix 
elements involving heavy quarks.  For a transition between
heavy ground-state mesons $H$  (either pseudoscalar or vector)
with heavy flavour $f$ ($f'$) moving with velocities $v$ ($v'$), one
obtains in the heavy-quark limit
\begin{equation} \label{WET}
\langle H^{(f')} (v') | \bar{h}^{(f')}_{v'} \Gamma h^{(f)}_v
|  H^{(f)} (v) \rangle 
 = \xi (vv') \mbox{ Tr } 
\left\{ \overline{{\cal H} (v)} \Gamma {\cal H}(v) \right\} ,
\end{equation}
where $\Gamma$ is some arbitrary Dirac matrix and $H(v)$ are the 
representation matrices for the two possibilities of coupling 
the heavy quark spin to the spin of the light degrees of freedom, 
which are in a spin-1/2 state for ground state mesons   
\begin{equation} \label{mesonrep}
{\cal H}(v) = \frac{\sqrt{M_H}}{2} \left\{ \begin{array}{l l}
        (1+\fmslash{v}) \gamma_5 & 0^-, \, (\bar{q} Q)  \mbox{ meson} \\
        (1+\fmslash{v}) \fmslash{\epsilon}  & 1^- , \, (\bar{q} Q) \mbox{ meson} \\
                            & \mbox{with polarization } \epsilon .
       \end{array} \right.  
\end{equation}  
Due to the spin and flavour independence of the heavy mass limit 
the Isgur--Wise function $\xi$ is the only non-perturbative information 
needed to describe all heavy to heavy transitions within a spin-flavour 
symmetry multiplet.  

Similar statements may be derived for the spin symmetry doublets of 
excited heavy mesons\cite{Falk} and also for 
baryons\cite{IWbary,Geobary,MRRbary}. 

\subsection{Corrections to the infinite mass limit}
Corrections to the infinite mass limit may be considered in a systematic way. 
They fall into two classes: The recoil or $1/m_Q$ corrections and the QCD
radiative corrections. 

In order to discuss the corrections we shall consider a specific example, 
namely the matrix element of a current 
$\bar{q} \Gamma Q$ mediating
a transition between a heavy meson and some arbitrary state 
$| A \rangle$. Using the expansions (\ref{HarvLag}) and (\ref{HarvField}) 
one obtains up to order $1/m_Q$ 
\begin{eqnarray} \label{exp}
&& \langle A | \bar{q} \Gamma Q 
              | M (v) \rangle =  
\langle A | \bar{q} \Gamma h_v 
              | H (v) \rangle  \\ \nonumber 
&& \qquad + \frac{1}{2m_Q} \langle A | \bar{q} \Gamma 
              P_- i \fmslash{D} h_v 
              | H (v) \rangle    
 -i \int d^4 x \langle A | T \{ L_1 (x) \bar{q} \Gamma h_v \}
              | H (v) \rangle   
   + {\cal O} (1/m^2)  
\end{eqnarray}
where $L_1$ are the first-order corrections to the Lagrangian
as given in (\ref{HarvLag}). Furthermore, $| M (v) \rangle $ is the 
state of the heavy meson in full QCD, including all its mass 
dependence, while $| H (v) \rangle$ is the corresponding state in 
the infinite mass limit.  

Expression (\ref{exp}) displays the generic structure of the 
higher-order corrections as they appear in any HQET calculation. 
There will be local contributions coming from the expansion of 
the full QCD field; these may be interpreted as the corrections to 
the currents. The non-local contributions, i.e.~the time-ordered
products, are the corresponding corrections to the states and thus
in the r.h.s.\ of (\ref{exp}) only the states of the infinite-mass 
limit appear.

Although the $1/m_Q$ corrections need in general additional input 
beyond HQET, there is one important result on the corrections 
linear in $1/m_Q$, which is called Luke's theorem\cite{Lu92} and
which is the application of the Ademollo Gatto theorem\cite{AGT} 
to the case of heavy flavour symmetry. In its general form the theorem 
states that in the presence of explicit symmetry breaking the 
matrix elements of the symmetry generating currents, which are 
normalized due to the symmetry, do not receive corrections linear
in the symmetry breaking. Applied to the case at hand this means 
that some of the form factors in weak decays, namely the ones
proportional to the Isgur Wise function, receive only corrections
quadratic in $1/m_Q$. This result has important phenomenological 
consequences, which we shall discuss below.    
  
All the relations given up to now are tree level relations. 
Going beyond tree level will induce QCD radiative corrections of  
order $\alpha_s^n (m_Q)$, $n = 1, ...$. As in any field theory 
these corrections are perturbatively calculable in terms of Feynman
diagramms. The effective theory has two additional Feynman rules 
(the propagator of the heavy quark and the coupling of the heavy 
quark to the gluons) which may be read off from the static Lagrangian 
(\ref{Lstat}).  

For the sake of clarity we shall stick to our example of a heavy light 
current considered above. To leading order in the $1/m_Q$ expansion one 
may evaluate the radiative corrections to such a matrix element using the 
above Feynman rules and finds a divergent result with a divergence related 
to the short distance behavior. Since HQET is an 
effective theory, the machinery of effective theory guarantees the 
factorization of long distance effects from the short distance ones, 
which are related to the large mass $m_Q$.  
Neglecting $1/m_Q$ corrections, this factorization takes the form
\begin{equation} \label{expr}
\langle A | \bar{q} \Gamma Q 
              | M (v) \rangle = Z \left( \frac{m_Q}{\mu} \right)  
\langle A | \bar{q} \Gamma h_v 
              | H (v) \rangle |_\mu 
   + {\cal O} (1/m_Q)  
\end{equation}
From Feynman rule calculation one obtains the perturbative expansion 
of the renormalization constant $Z$ which generically looks like  
\begin{eqnarray} \label{zexp}
Z \left( \frac{m_Q}{\mu} \right) &=& a_{00} \\
&+& a_{11} \left( \alpha_s \ln \left(\frac{m_Q}{\mu} \right) \right)\,
 +  a_{10} \alpha_s \nonumber \\ 
&+& a_{22} \left( \alpha_s \ln \left(\frac{m_Q}{\mu} \right) \right)^2
 +  a_{21} \alpha_s
           \left( \alpha_s \ln \left(\frac{m_Q}{\mu} \right) \right)
 +  a_{20} \alpha_s^2 \nonumber \\
&+& a_{33} \left( \alpha_s \ln \left(\frac{m_Q}{\mu} \right) \right)^3
 +  a_{32} \alpha_s
           \left( \alpha_s \ln \left(\frac{m_Q}{\mu} \right) \right)^2
 +  a_{31} \alpha_s^2
           \left( \alpha_s \ln \left(\frac{m_Q}{\mu} \right) \right) 
\nonumber \\
&& \hphantom{--------} +  a_{30} \alpha_s^3 
 + \cdots \nonumber
\end{eqnarray}
where $\alpha_s = g^2/(4\pi)$.

This factorization theorem corresponds to the statement that the
ultraviolet divergencies in the effective theory have to match the
logarithmic mass dependences of full QCD. The factorization scale $\mu$ is
an arbitrary parameter, and the physical quantity
$\langle A | \bar{q} \Gamma Q | M (v) \rangle $
does not depend on this parameter. However, calculating
the matrix element of this operator in the effective theory and
studying its ultraviolet behavior allows us to access the mass dependence
of the matrix element $\langle A | \bar{q} \Gamma Q | M (v) \rangle $.

The ultraviolet behavior of the effective theory is investigated by
the renormalization group equation (RGE), which is obtained in the usual way 
from differentiating (\ref{expr}) with respect to the factorization 
scale $\mu$. The RGE for the short distance coefficient $Z$ becomes
\begin{equation} \label{rge}
\left( \mu \frac{\partial}{\partial \mu}
+ \beta(g) \frac{\partial}{\partial g} +
\gamma_J (g) \right)  Z \left( \frac{m_Q}{\mu}, g \right) = 0 .
\end{equation}
where $\gamma_J (g)$ is the anomalous dimension of the current $J$ 
which is related to the ultraviolet behavior of the matrix elements of 
$J$. The function $\beta(g)$ defines the running of the coupling constant 
\begin{equation} \label{gml}
\frac{d}{d \ln \mu} g (\mu) = \beta (\mu) .
\end{equation}
Both functions $\gamma_J (g)$ and $\beta(g)$ are calculable
in perturbation theory using a loopwise expansion, where the first term 
of the $\beta$ function of QCD is well known
\begin{equation} \label{beta0}
\beta(g) = - \frac{1}{(4 \pi)^2} \left( 11 - \frac{2}{3} n_f \right)
           g^3 + \cdots ,
\end{equation}
where $n_f$ is the number of flavors with a mass less than $m_Q$.

With this input the renormalization group equation may be solved to
yield
\begin{equation} \label{LLA}
Z \left( \frac{m_Q}{\mu} \right) = a_{00} \left( 
\frac{\alpha_s (\mu)}{\alpha_s (m_Q)} \right)^
          {\displaystyle -\frac{48 \pi^2}{33 - 2 n_f} \gamma_1}
\end{equation}
where $\gamma_1$ is the first coefficient in the perturbative expansion of
the anomalous dimension $\gamma_J = \gamma_1 g^2 + \cdots $ and 
$\alpha_s (\mu)$ is the one loop expression for the running coupling
constant of QCD
\begin{equation} \label{alpha}
\alpha_s (\mu) = \frac{12 \pi}{(33-2n_f) \ln(\mu^2/\Lambda_{QCD}^2)}
\end{equation}
which is obtained from solving (\ref{gml}) using (\ref{beta0}).
 
This expression corresponds to a summation of the leading logarithms
$(\alpha_s \ln m_Q)^n$ which is  achieved by a one-loop calculation 
of the renormalization group functions $\beta$ and $\gamma_{\cal Q}$;  
in other words, in this way a resummation of the
first column of the expansion (\ref{zexp}) is achieved.

In a similar way one may also resum the second column of (\ref{zexp}), 
if the renormalization group functions $\beta$ and $\gamma$ are calculated 
to two loops and the non-logarithmic terms of the one loop 
expression are included.

Eq.(\ref{rge}) describes the renormalization group scaling
in the effective theory. It allows to shift logarithms of the large
mass scale from the matrix element of $J$ into the 
coefficient $Z$: If the matrix element is renormalized at the large
scale $m_Q$ the logarithms of the type $\ln m_Q$ will 
apear in the matrix element
of $J$ while the coefficient $Z$ at this scale will simply be 
\begin{equation}
Z(1) = a_{00} + a_{10} \alpha_s (m_Q) + a_{20} \alpha_s^2 (m_Q) 
              + a_{30} \alpha_s^3 (m_Q) + \cdots
\end{equation}
The renormalization
group equation (\ref{rge}) allows to lower the renormalization point
from $m_Q$ to $\mu$; the matrix element renormalized at $\mu$ will
not contain any logarithms of $m_Q$ any more, they will appear in
the coefficient $Z$ in the way shown in (\ref{zexp}). 

In all cases relevant in the present context the matrix elements will 
be matrix elements involving
hadronic states, which are in most cases impossible to calculate from
first principles. However, eq.(\ref{rge}) allows to extract the
short distance piece, i.e.\ the logarithms of the large mass $m_Q$
and to separate it into the Wilson coefficients.

Finally, the case we have considered as an example is indeed very simple; 
in general all operators of a given dimension may mix under renormalization, 
i.e. instead of a simple anomalous dimension a matrix of anomalous dimensions
may occur and the renormalization group equation \ref{rge} becomes a 
system of differential equations.  

\boldmath
\section{Exclusive semileptonic $b\to c$ transitions}
\unboldmath
In this section we shall discuss the transitions of the heavy to heavy 
type, i.e. the $b \to c$ decays. The implications of heavy quark 
symmetry have been given already in the form of the Wigner Eckart theorem 
(\ref{WET}) in the last section, so we shall focus here on the status of 
the corrections to the infinite mass limit and give the phenomenological 
applications of the results.

\subsection{QCD Radiative Corrections}
The matrix elements relevant for the processes under consideration are 
\begin{equation} \label{currents}
V_\mu = \langle D^{(*)} (v') | \bar{c} \gamma_\mu b | B(v) \rangle ,
\qquad
A_\mu = \langle D^* (v') | \bar{c} \gamma_5 \gamma_\mu b | B(v) \rangle
\end{equation}
where the point $v = v'$ is of particular interest, since the absolute 
normalization of the matrix elements is known at this point 
due to heavy quark symmetries.

In order to evaluate the QCD radiative corrections to these matrix elements
we shall make use of the renormalization group (RG) machinery as outlined 
in the last section. 
At a large scale $\mu \sim M_W$ both currents are conserved in the limit 
of vanishing $b$ and $c$ quark masses and hence their anomalous 
dimension vanishes. Running down from $M_W$ to $m_b$, i.e. lowering the 
renormalization scale of the matrix elements (\ref{currents}) from $M_W$
to $m_b$ induces no large logarithms of the form 
$\alpha_s (M_W) \ln (M_W / m_b)$, rather the corrections will be small, 
of the order $\alpha_s (M_W) / \pi$. 

Similarly, at scales $\mu$ below the charm quark mass $m_c$ both quarks 
may be taken to be infinitely heavy, and at the non-recoil point $v = v'$
again the two currents are conserved and thus their anomalous dimension 
vanishes. Running below the charm quark mass will thus induce only small 
corrections of the order $\alpha_s (m_c)$. 

Thus the main corrections originate from scales $\mu$ between $m_b$ and 
$m_c$. In the effective theory where the $b$ is taken to be infinitely 
heavy and the $c$ is still light the one\cite{oneloop} and two 
loop\cite{twoloop} anomalous dimensions have been calculated and    
allow a resummation of terms of order  $(\alpha_s (m_b) \ln (m_b / m_c))^n$, 
and $\alpha_s (m_b)(\alpha_s (m_b) \ln (m_b / m_c))^{n-1}$ respectively. 
Furthermore, in this theory the subleading terms of order $1/m_b$ 
\cite{fg92} and $1/m_b^2$ \cite{bo96} have been considered at the one 
loop level, and the matching at the scale $m_c$ yields a resummation 
of terms of order $(m_b / m_c)  (\alpha_s (m_b) \ln (m_b / m_c))^n$
and $(m_b / m_c)^2  (\alpha_s (m_b) \ln (m_b / m_c))^n$. 

The procedure described here has the disadvantage that numerically it 
is useful only in the limit $m_b \gg m_c$ such that $\ln (m_b / m_c) \gg 1$
is a large logarithm and terms of order 
$(m_c / m_b)^n  \alpha_s (m_b) \ln (m_b / m_c))$ may be neglected.
In real life we have $m_b /m_c \sim 3$ and one may 
think of simply performing the one loop calculation including the masses
in full QCD and use this to match directly QCD to a theory with two 
static quarks, where heavy quark symmetries hold. In this way one 
obtains all terms of order $(m_c / m_b)^n  \alpha_s  \ln (m_b / m_c))$. 
However, the price to pay is a scale ambiguity in the scale of $\alpha_s$
which has to be taken at some scale $\bar{m}$ between $m_b$ and $m_c$. 
Numerically this is not a problem since $\alpha_s$ does not run much between 
$m_c$ and $m_b$. 

Recently the latter procedure has ben applied at the two loop level 
\cite{Czarn} such that the terms of order 
$(m_b / m_c)^n  \alpha_s^2 (\bar{m})  \ln (m_b / m_c))$ are now known. 
The usual way to parametrize the corrections is by introducing 
corrections factors 
\begin{equation}
V_\mu \to \eta_V V_\mu , \qquad  A_\mu \to \eta_A A_\mu
\end{equation}
which are known up to terms of order $\alpha_s^3 (\bar{m})$. Taking 
the numbers of \cite{Czarn} one has 
\begin{eqnarray}
\eta_V &=& 1 + 0.018 + 0.004 + {\cal O}(\alpha_s^3 (\bar{m})) 
        =  1.022 \pm 0.004 \\
\eta_A &=& 1 - 0.033 - 0.007 + {\cal O}(\alpha_s^3 (\bar{m})) 
        =  0.960 \pm 0.007
\end{eqnarray}
The uncertainty given here is the uncertainty due to the terms of order
$\alpha_s^3 (\bar{m})$ and is conservatively estimated by the size of 
the calculated $\alpha_s^2 (\bar{m})$ corrections. 

Given the fact that the calculated QCD corrections are already below the
level of $1 \%$ one needs to worry also about QED corrections. In fact, 
the QED corrections to processes of this type have been calculated 
already some time ago in the context of $\mu$ decay\cite{sirlin}. 
Similar to the 
QCD corrections they may be factorized and in principle one could apply 
also renormalization group methods. However, the running of the QED
coupling is negligible at this level and the QED corrections factor 
is given by
\begin{equation}
\eta_{A/V} \to  \eta_{A/V} \left[1 + \frac{\alpha_{QED}}{\pi}
                           \ln \frac{M_Z}{m_b} \right] 
           \sim 1.013 \,\, \eta_{A/V}
\end{equation}
Hence short distance QED corrections enhance the matrix elements by 
$1.3 \%$.

\subsection{Recoil Correction}
In general the recoil corrections are more complicated, since they may
not be calculated from the effective theory, rather they can only be 
parametrized in terms of new matrix elements. The calculation of these matrix
elements in any case needs additional input (such as models or a 
lattice calculation) beyond the framework of HQET. 

However, in some cases Luke's theorem applies.
In particular, the axial vector current taken 
at the non-recoil point $v = v'$ is protected by this  theorem, such 
that 
\begin{equation}
\langle D^* (v') | \bar{c} \gamma_5 \gamma_\mu b | B(v) \rangle
= 2 \sqrt{M_{D^*} M_B} \epsilon_\mu 
    \left(1 + \delta_{1/m^2} \right), \quad 
\delta_{1/m^2} = {\cal O} \left(\frac{1}{m_b^2}, \, \frac{1}{m_c^2}
                             \frac{1}{m_c m_b} \right) 
\end{equation}
Among the first non-trivial corrections the ones of the order
$1/m_c^2$ will be the most important, and HQET allows only to parametrize
them in terms of new matrix elements. One obtains
\begin{eqnarray} \label{oom2}
\delta_{1/m^2}  &=&  - \left(\frac{1}{2m_c} \right)^2
   \frac{1}{2} \left( -\lambda_1 + \lambda_2  \vphantom{\int} \right.
\\ \nonumber  &+&   \left. 
(-i)^2 \frac{1}{2\sqrt{M_B M_D}} \int d^4 x \,d^4y \,
\langle B^*(v,\epsilon) | T \left[ {\cal L}_b^{(1)} (x) \bar{b}_v c_v 
                                      {\cal L}_c^{(1)} (y) \right]
                   | D^*(v,\epsilon) \rangle \right) \\
&+& \vphantom{\int}{\cal O} (1/m_c^3,1/m_b^2, 1/(m_c m_b) ) ,  \nonumber
\end{eqnarray}
where ${\cal L}_Q^{(1)}$ is the first order Lagangian for 
the quark $Q$ as given  in (\ref{HarvLag}). Furthermore, 
the parameters $\lambda_1$ and $\lambda_2$ are the kinetic energy and 
the chromomagnetic moment of the heavy quark. They also appear in the 
$1/m_Q$ expansion of the heavy meson mass
\begin{equation} \label{massrel}
m_H = m_Q \left(1 + \frac{\bar\Lambda}{m_Q} + \frac{1}{2 m_Q^2} 
      \left(\lambda_1 + d_H \lambda_2 \right)
+ {\cal O} (1/m_Q^3) \right)  
\end{equation}
where $d_H = 3$ for the $0^-$ and $d_H = -1$ for the $1^-$ meson, and 
may be related to the matrix elements 
\begin{eqnarray} 
\bar{\Lambda} &=& 
\frac{\langle 0 | q \stackrel{\longleftarrow}{ivD} \gamma_5 h_v 
     | H (v) \rangle}
     {\langle 0 | q \gamma_5 h_v | H (v) \rangle}
\label{Lambar} \\ 
\lambda_1 &=& \frac{\langle H (v) | \bar{h}_v  (iD)^2  h_v | H (v) \rangle}
                   {2 M_H} 
\label{lam1} \\
\lambda_2  &=& \frac{\langle H (v) | 
                            \bar{h}_v \sigma_{\mu \nu} iD^\mu iD^\nu h_v 
                                   | H (v) \rangle}{2 M_H}
\label{lam2}
\end{eqnarray}
The only parameter which is easy to access is
$\lambda_2$, since it is related to the mass splitting between $H(v)$ 
and $H^* (v , \epsilon)$. From the $B$-meson system we obtain
\begin{equation}
\lambda_2 (m_b) = \frac{1}{4} (M_{H^*} - M_H) = 0.12 \mbox{ GeV}^2; 
\end{equation}
and from the charm system (including the renormalization group scaling of 
$\lambda_1$) the same value is obtained.  This shows that indeed
the spin-symmetry partners are degenerate in the infinite mass limit and 
the splitting between them scales as $1/m_Q$. 

The other parameters appearing in (\ref{massrel}) are not simply 
related to the hadron spectrum. Furthermore, they exhibit renormalon 
ambiguities, which imply that a proper prescription has to be given 
how to extract these quantities from data using renormalized perturbation 
theory. 

Recently there has been such an attempt \cite{GKLW96}, namley to extract 
$\bar\Lambda$ and $\lambda_1$ from the shape of the lepton energy spectrum 
in inclusive semileptonic 
$B$ decays. The values obtained from this analysis are
$\bar\Lambda = 0.39 \pm 0.11$ GeV and $-\lambda_1 = 0.19 \pm 0.10$ GeV${}^2$, 
where the $\overline{\rm MS}$ definition of the mass has been used. 
The uncertainties
quoted are only the $1\sigma$ statistical ones; the systematical uncertainties 
of this approach are difficult to estimate.  
 
These considerations only fix the local matrix elements in (\ref{oom2}), 
while the nonlocal terms involving the time-ordered products are much
harder to estimate. The estimates found in the literature have to 
rely on some model estimates and span a range which implies roughly 
a $3 \%$ theoretical uncertainty. A commonly accepted number was given
in \cite{oom2Neu}
\begin{equation}
\delta_{1/m^2} = - 0.0055 \pm 0.025  
\end{equation}

\subsection{Phenomenology}
\subsubsection{Differential rates}
In the last few years a lot of data has been accumulated which may be used 
to test heavy quark symmetry and to extract CKM matrix elements with the 
use of HQET. The exclusive decays of prime interest are the transitions 
$B \to D \ell \nu_\ell$ and $B \to D^* \ell \nu_\ell$. The relevant 
matrix elements are
\begin{eqnarray}
 \langle D (v') | \bar{c} \gamma_\mu b | B(v) \rangle  &=&  \sqrt{m_B m_D}
 \left[ \xi_+ (y) (v_\mu + v'_\mu)
     + \xi_- (y) (v_\mu - v'_\mu) \right] \\ 
 \langle D^* (v',\epsilon) | \bar{c} \gamma_\mu b | B(v) \rangle  &=& 
       i \sqrt{m_B m_{D^*}} 
\xi_V (y) \varepsilon_{\mu \alpha \beta \rho} \epsilon^{*\alpha}
                       v^{\prime \beta} v^\rho  \\ 
\langle D^* (v',\epsilon) | \bar{c} \gamma_\mu \gamma_5 b | B(v) \rangle 
   &=& \sqrt{m_B m_{D^*}}
       \left[ \xi_{A1} (y) (vv'+1) \epsilon^*_\mu
      -  \xi_{A2} (y) (\epsilon^* v)  v_\mu \right. \nonumber \\
   && \qquad \qquad \left.   -  \xi_{A2} (y) (\epsilon^* v)  v'_\mu \right] ,
\end{eqnarray} 
where we have defined $y = vv'$. Thus in general there are six form factors,
which in the heavymass limit for both the $b$ and the $c$ quark may be 
related to the Isgur Wise function as introduced in (\ref{WET})
\begin{equation}
\xi_i (y) = \xi (y) \mbox{ for } i = +,V,A1,A3, \qquad
\xi_i (y) = 0       \mbox{ for } i = -,A2 .
\end{equation} 
In particular, at the non-recoil point $v=v'$ we have due to heavy quark 
symmetry and Lukes theorem
\begin{equation}
\xi_i (1) = 1 + {\cal O} (1/m_Q^2) \mbox{ for } i = +,V,A1,A3, \quad
\xi_i (1) = {\cal O} (1/m_Q)     \mbox{ for } i = -,A2 .
\end{equation} 
Note that the form factors for which there is no normalization at 
$v=v'$ there is also no protection against corrections of lienar 
order in $1/m_Q$.  

The differential rates for the exclusive semiletonic $b \to c$ 
transitions may be expressed in terms of the six form factors as
\begin{eqnarray}
&& \frac{d\Gamma}{dy} (B \to D \ell \nu_\ell)  \label{bd} \\
&& = 
\frac{G_F^2}{48 \pi^3} |V_{cb}|^2 (m_B + m_D)^2 
\left(m_D \sqrt{y^2 -1}\right)^3 
\left| \xi_+ (y) - \frac{m_B - m_D}{m_B + m_D} \xi_- (y) \right|^2 
\nonumber \\
&&\frac{d\Gamma}{dy} (B \to D^* \ell \nu_\ell) \label{bds}   \\
&& = \frac{G_F^2}{48 \pi^3} |V_{cb}|^2 (m_B - m_{D^*})^2 m_{D^*}^2
\left(m_{D^*} \sqrt{y^2 -1}\right) (y+1)^2 |\xi_{A1} (y)|^2 
\sum_{i=0,\pm} |H_i(y)|^2 \nonumber 
\end{eqnarray} 
with 
\begin{eqnarray}
|H_\pm (y)|^2 &=& \frac{m_B^2 - m_{D^*}^2 - 2y m_B m_{D^*}}{(m_B - m_{D^*})^2}
\left[1 \mp \sqrt{\frac{y-1}{y+1}} R_1 (y) \right]^2 \\
|H_0 (y)|^2 &=& \left(1 + \frac{m_B (y-1)}{m_B - m_{D^*}} 
                \left[1-R_2 (y) \right] \right)^2
\end{eqnarray}
where we have defined the form factor ratios 
\begin{equation}
R_1 (y) = \frac{\xi_V (y)}{\xi_{A1} (y)}, \quad 
R_2 (y) = \frac{\xi_{A3} (y) + \frac{m_B}{m_{D^*}} \xi_{A2} (y)}{\xi_{A1} (y)}
\end{equation}
In the heavy mass limit these differential rates depend only on the 
Isgur-Wise function
\begin{eqnarray}
&& \frac{d\Gamma}{dy} (B \to D \ell \nu_\ell)  
\stackrel{m_b, m_c \to \infty}{=} 
\frac{G_F^2}{48 \pi^3} |V_{cb}|^2 (m_B + m_D)^2 
\left(m_D \sqrt{y^2 -1}\right)^3 
| \xi (y)|^2 \\
&& \frac{d\Gamma}{dy} (B \to D \ell \nu_\ell)   
\stackrel{m_b, m_c \to \infty}{=} \\
&&  \frac{G_F^2}{48 \pi^3} |V_{cb}|^2 (m_B - m_{D^*})^2 m_{D^*}^2
\left(m_{D^*} \sqrt{y^2 -1}\right) (y+1)^2 \\ \nonumber 
&& \qquad \left[ 1 + \frac{4y}{y+1}
\frac{m_B^2 - m_{D^*}^2 - 2y m_B m_{D^*}}{(m_B - m_{D^*})^2} \right]
| \xi (y)|^2 \nonumber 
\end{eqnarray} 
These relations allow for a test of heavy quark symmetry, since the 
ratios of the differential rates do not depend on any unknown form factor 
any more. In particular the ratios $R_1$ ($R_2$) measures the ratio of 
the differential transverse (longitudinal) rate and the total differential 
rate. In the heavy mass limit both $R_1$ and $R_2$ are unity; this has to 
be compared to the measurements by CLEO \cite{CLEORi,cassel}
\begin{eqnarray}
R_1 &=& 1.24 \pm 0.26 \pm 0.12 \\
R_2 &=& 0.72 \pm 0.18 \pm 0.07 
\end{eqnarray}

\subsubsection{The determination of $V_{cb}$}
From the measured lepton invariant mass spectrum one may determine $V_{cb}$
in a model independent way by extrapolating to the kinematical 
endpoint of maximal momentum transfer to the leptons, corresponding
to the point $v = v'$. At this point heavy quark symmetries determine the 
absolute normalization of some of the form factors and the corrections 
to this normalization have been discussed in section 2.

The mode $B \to D^* \ell \nu_\ell$ has the advantage of a higher 
branching fraction and hence we shall start the discussion with 
this decay.
The relevant formula may be derived from (\ref{bds}) and reads
\begin{equation}
\lim_{y \to 1} \frac{1}{\sqrt{y^2 - 1}} \frac{d\Gamma}{dy}
(B \to D^* \ell \nu_\ell) = \frac{G_F^2}{4 \pi^3} (m_B - m_{D^*})^2 
m_{D^*}^3 |V_{cb}|^2 |\xi_{A1} (1)|^2 
\end{equation}
The form factor $\xi_{A1}$ is normalized due to heavy quark symmetries
and is hence protected against $1/m_Q$ corrections at $v = v'$ by Lukes 
theorem. Hence we have
\begin{equation}
\xi_{A1} (1) = \eta_A (1+\delta_{1/m^2}) 
\end{equation}
Including QED corrections and the estimate of the $1/m_Q^2$ corrections 
in the way discussed in section 2 one obtains
\begin{equation}
\xi_{A1} (1) = 0.92 \pm 0.03
\end{equation}
From this value one may extract from the extrapolations shown 
in figs.1 and 2
a value for $V_{cb}$
\begin{eqnarray}
|V_{cb}| &=& 0.0362 \pm 0.0019 \pm 0.0020 \pm 0.0014 
             \mbox{ (CLEO)\cite{cleovcb}} \\
|V_{cb}| &=& 0.0345 \pm 0.0025 \pm 0.0027 \pm 0.0015 
             \mbox{ (ALEPH)\cite{ALEPHvcb}} 
\end{eqnarray}
where the last error reflects the theoretical uncertainty of 
the $1/m_Q$ corrections. 
%

Recently CLEO has also measured the leptonic invariant mass spectrum 
for the decay $B \to D \ell \nu_\ell$. In a similar way one may perform 
an extrapolation to obtain $V_{cb}$. Here one gets from (\ref{bd})
\begin{eqnarray}
\lim_{y \to 1} && \left(\frac{1}{\sqrt{y^2 - 1}}\right) \frac{d\Gamma}{dy}
(B \to D \ell \nu_\ell) 
\\ \nonumber && = \frac{G_F^2}{48 \pi^3} (m_B + m_D)^2 
m_D^3 |V_{cb}|^2 
\left| \xi_+ (1) - \frac{m_B - m_D}{m_B + m_D} \xi_- (1) \right|^2 
\end{eqnarray}
In this case a form factor enters which is not protected by 
Lukes theorem, $\xi_- (1) = {\cal O} (1/m_Q)$. However, this does not 
spoil the possibility to determine $|V_{cb}|$ from this mode, since the 
$1/m_Q$ corrections are kinematically suppressed by the 
factor $ (m_B - m_D)/(m_B + m_D)$. Here we have 
\begin{equation}
\left| \xi_+ (1) - \frac{m_B - m_D}{m_B + m_D} \xi_- (1) \right| = 
\eta_V (1+ \Delta_{1/m_Q})
\end{equation}
where $\Delta_{1/m_Q}$ are the $1/m_Q$ corrections induced by $\xi_- (1)$.
These corrections have been estimated recently \cite{CapNeu} 
\begin{equation} \label{bdest}
\left| \xi_+ (1) - \frac{m_B - m_D}{m_B + m_D} \xi_- (1) \right| = 
0.98 \pm 0.07
\end{equation}
The data from CLEO{cleovcb1} are shown in fig.3. 
From the extrapolation 
shown in fig.\ref{fig2} one obtain the value 
\begin{equation}
|V_{cb}| = 0.0353 \pm 0.0046 \pm 0.0044 \pm 0.0025
\end{equation}
where the last error again reflects the theoretical uncertainty. 
Both values of $V_{cb}$ are in very good agreement and are also consistent
with values obtained form other methods, in particular with the 
results from inclusive semileptonic decays\cite{Bigi}.

\subsubsection{The slope of the form factors}
The data on the leptonic invariant mass spectrum extend over the 
whole rage of $y$ and the extrapolation to $y = 1$ has to rely 
on some ansatz for the Isgur Wise function. This may be used in 
turn to extract a value also for the slope of the Isgur Wise function.  

Close to the point $y=1$ one thus parametrizes the Isgur Wise function 
as
\begin{equation}
\xi (y) = 1 - \rho^2 (y-1) + \cdots
\end{equation}
The theoretical predictions\cite{voloshin}
for the slope parameter $\rho^2$ 
depend on matrix elements 
involving excited $D$ mesons which are hard to estimate; thus the
theoretical value is quite uncertain and ranges between 
\begin{equation}
0.5 \le \rho^2 \le 1.1
\end{equation}
Data on the slope parameter come from the LEP experiments as well 
as from ARGUS and CLEO, the results obtained from experiment are 
\begin{equation}
\rho^2 = \left\{ \begin{array}{cc}
                 0.29 \pm 0.21 \pm 0.12 & \mbox{ALEPH\cite{alephrho}} \\
                 0.84 \pm 0.12 \pm 0.08 & \mbox{CLEO\cite{cleorho}} \\
                 1.17 \pm 0.22 \pm 0.06 & \mbox{ARGUS\cite{argusrho}} \\
                 0.81 \pm 0.16 \pm 0.10 & \mbox{DELPHI\cite{delphirho}}
                 \end{array} \right.
\end{equation}
which are compatible with the theoretical expectation.

\boldmath
\section{Exclusive leptonic and semileptonic $b\to u$ transitions}
\unboldmath
Heavy quark symmetries may also be used to restrict the independent
form factors appearing in heavy to light decays. For the decays of heavy 
mesons into light $0^-$ and $1^-$ particles heavy quark symmetries restrict 
the number of independent form factors to six, which is just the 
number needed to parametrize the semileptonic decays of this type. 
Furthermore, no absolute normalization of form factors may be obtained from 
heavy quark symmetries in the heavy to light case; only the relative 
normalization of $B$ meson decays heavy to light transitions may be obtained 
from the corresponding $D$ decays.

In general we shall discuss matrix elements of a heavy to light
current which have the following structure
\begin{equation} \label{htlg}
J = \langle  A | \bar{q} \Gamma h_v | H (v) \rangle ,
\end{equation}
where $\Gamma$ is an arbitrary Dirac matrix,
$q$ is a light quark
($u$, $d$ or $s$) and
$A$ is a state involving only light degrees of freedom.

Spin symmetry implies that the heavy quark index hooks directly 
to the heavy quark index of the Dirac matrix of the current. Thus one
may write for the transition matrix element~(\ref{htlps})
\begin{equation} \label{htltrg}
\langle  A | \bar{q} \Gamma h_v | H (v) \rangle =
\mbox{ Tr } \left( {\cal M}_A \Gamma H(v) \right)
\end{equation}
where the matrix $H(v)$ representing the heavy meson has been
given in (\ref{mesonrep}).
The matrix ${\cal M}_A$ describes the light degrees of freedom
and is the most general matrix which may
be formed from the kinematical variables involved. Furthermore, if
the energies of the particles in the state $A$ are small, i.e.\ of
the order of $\Lambda_{QCD}$, the matrix ${\cal M}_A$ does not depend
on the heavy quark; in particular it does not depend on the heavy mass
$m_H$.
In the following we
shall discuss some examples.

The first example is the heavy meson decay constant, where the
state $A$ is simply the vacuum state. The heavy meson decay
constant is defined by
\begin{equation}
\langle 0 | \bar{q} \gamma_\mu \gamma_5 h_v | H (v) \rangle =
f_H m_H v_\mu , 
\end{equation}
and since $| A \rangle = | 0 \rangle$ the matrix ${\cal M}_0$ is simply the 
unit matrix times a dimensionful constant\footnote{%
  Note that contributions proportional to
  $\fmslash{v}$ may be eliminated using
  $$
  H(v) \fmslash{v} = - H(v).
  $$ }
and one has, using
(\ref{htltrg})
\begin{equation}
\langle 0 | \bar{q} \gamma_\mu \gamma_5 h_v | H (v) \rangle = \kappa
\mbox{ Tr } \left( \gamma \gamma_5 H(v) \right) = 2 \kappa
\sqrt{m_H} v_\mu .
\end{equation}
As discussed above the constant $\kappa$ does not depend on the
heavy mass and thus one infers the well-known
scaling law for the heavy meson
decay constant from the last two equations
\begin{equation} \label{sca}
f_H  \propto \frac{1}{\sqrt{m_H}}
\end{equation}
Including the leading and subleading QCD radiative corrections
one obtains a relation between $f_B$ and $f_D$
\begin{equation}
f_B  = \sqrt{ \frac{m_c}{m_b}}
\left( \frac{\alpha_s (m_b)}{\alpha_s (m_c)} \right)^{-6/25} 
\left[ 1 + 0.894 \frac{\alpha_s (m_c)-\alpha_s (m_b)}{\pi} \right]
f_D \sim 0.69 f_D .
\end{equation}

The second example are transitions of a heavy meson
into a light pseudoscalar meson, which we shall denote as $\pi$.
The matrix element corresponding to (\ref{htlg}) is
\begin{equation} \label{htlps}
J_P = \langle \pi(p) | \bar{q} \Gamma h_v | H (v) \rangle , 
\end{equation}
where $p$ is the momentum of the light quark,

The Dirac marix ${\cal M}_P$ for the light degrees of freedom
appearing now in (\ref{htltrg})
depends on $p$ and $v$.
It may be expanded in terms of the sixteen independent
Dirac matrices $1$, $\gamma_5$,  $\gamma_\mu$, $\gamma_5 \gamma_\mu$,
and $\sigma_{\mu \nu}$ taking into account that it has to behave
like a pseudoscalar.
The form factors appearing in the decomposition of
${\cal M}_P$
depend on the variable $v \cdot p$, the energy of the light meson
in the rest frame of the heavy one. In order to compare different
heavy to light transition by employing heavy flavor symmetry this
energy must be sufficiently small, since the typical scale for the
light degrees of freedom has to be of the order of $\Lambda_{QCD}$
to apply heavy quark symmetry\footnote{%
         Note that in this case the variable $v \cdot p$ ranges between
         $0$ and $m_H / 2$ where we have neglected the pion mass. Thus 
         at the upper end of phase space the variable $v \cdot p$ scales
         with the heavy mass and heavy quark symmetries are not applicable
         any more.}.  
For the case of a light pseudoscalar
meson the most general decomposition of ${\cal M}_P$ is
\begin{equation}
{\cal M}_P = \sqrt{v \cdot p} A (\eta ) \gamma_5
         + \frac{1}{\sqrt{v \cdot p}} B (\eta ) \gamma_5 \fmslash{p} ,
\end{equation}
where we have defined the dimensionless variable
\begin{equation} \label{eta}
\eta = \frac{v \cdot p}{\Lambda_{QCD}} .
\end{equation}

The form factors $A$ and $B$ are universal in the kinematic range
of small energy of the light meson, i.e.\ where the momentum
transfer to the light degrees of freedom is of the order $\Lambda_{QCD}$;
in this region $\eta$ is of order unity. This universality of the
form factors may be used to relate various kinds of heavy to
light transitions, e.g.\ the semileptonic decays like
$D \to \pi e \nu$, $D \to K e \nu$ or $B \to \pi e \nu$
and also the rare decays like $B \to K \ell^+ \ell^-$
or $B \to \pi \ell^+ \ell^-$ where $\ell$ denotes an electron or a muon.

As an example we give the relations between exclusive semileptonic
heavy to light decays. The relevant hadronic
current for this case may be expressed in terms of two form
factors
\begin{eqnarray} \label{frmfps}
\langle \pi(p) | \bar{q} \gamma (1-\gamma_5) h_v | H (v) \rangle
&=& F_1 (v \cdot p ) m_H v_\mu +
    F_2 (v \cdot p ) p_\mu
\\
&=& F_+ (v \cdot p ) (m_H v_\mu + p_\mu ) +
    F_- (v \cdot p ) q_\mu
\nonumber
\end{eqnarray}
where
\begin{equation}
F_\pm (v \cdot p ) = \frac{1}{2} \left( F_1 (v \cdot p ) \pm
                     F_2 (v \cdot p ) \right)
\end{equation}
Inserting this into (\ref{htlps}) one may express $F_\pm$ in terms
of the universal form factors $A$ and $B$
\begin{eqnarray} \label{f1hql}
F_1 (v \cdot p ) &=& F_+ (v \cdot p ) + F_- (v \cdot p )
=   -2 \sqrt{\frac{v \cdot p}{m_H}} A (\eta)
\\  \label{f2hql}
F_2 (v \cdot p ) &=& F_+ (v \cdot p ) - F_- (v \cdot p )
=   -2 \sqrt{\frac{m_H}{v \cdot p}} B (\eta)
\end{eqnarray}
From these relations one may read off the scaling of the form
factors with the heavy mass which was already
derived in \cite{iwlight}.

This may be used to normalize the semileptonic $B$ decays into light mesons
relative to the semileptonic $D$ decays. One obtains
\begin{equation} \label{bdphql}
F_\pm^B (v \cdot p ) =
\frac{1}{2} \left(\sqrt{\frac{m_D}{m_B}}
                  \pm \sqrt{\frac{m_B}{m_D}} \right)
                       F_+^D (v \cdot p )
+ \frac{1}{2} \left(\sqrt{\frac{m_D}{m_B}}
                  \mp \sqrt{\frac{m_B}{m_D}} \right)
                       F_-^D (v \cdot p )
\end{equation}
Note that $F_+$ for the $B$ decay is expressed in terms of
$F_+$ {\it and} $F_-$ for the $D$ decays. In the limit of vanishing
fermion masses only $F_+$ contributes, which means that the $F_-$ 
contribution to the rate is of the
order of $m_{lepton} / m_H$. Thus it will be extremely difficult to
determine experimentally.

The case of a heavy meson decaying into a light vector meson may be
treated similarly.
The matrix element for the transition of a heavy meson into a light
vector meson (denoted generically as $\rho$ in the following)
is given again by (\ref{htlg}) and is in this case
\begin{equation} \label{htlv}
J_V = \langle \rho(p,\epsilon) | \bar{q} \Gamma h_v | H (v) \rangle .
\end{equation}
Using (\ref{htltrg}) one has
\begin{equation} \label{htlvtr}
\langle \rho(p,\epsilon) | \bar{q} \Gamma h_v | H (v) \rangle =
\mbox{ Tr } \left( {\cal M}_V \Gamma H(v) \right) ,
\end{equation}
where now the Dirac matrix ${\cal M}_V$ has to be a linear
function of the polarization of the light vector meson.

The most general decomposition
is given in terms of four
dimensionless form factors
\begin{equation} \label{hqlv}
{\cal M}_V = \sqrt{v \cdot p} C(\eta) (v \cdot \epsilon)
    +  \frac{1}{\sqrt{v \cdot p}} D(\eta) (v \cdot \epsilon) \fmslash{p}
 + \sqrt{v \cdot p} E(\eta) \fmslash{\epsilon}
 +  \frac{1}{\sqrt{v \cdot p}} F(\eta) \fmslash{p} \fmslash{\epsilon}
\end{equation}
where the variable $\eta$ has been defined in (\ref{eta}).

Similar to the case of the decays into a light pseudoscalar meson
(\ref{htlvtr}) may be used to relate various exclusive heavy to
light processes in the kinematic range where the energy of the
outgoing vector meson is small. For example,
the semileptonic decays
$D \to \rho e \nu$, $D \to K^* e \nu$ and $B \to \rho e \nu$ are related
among themselves and all of them may be related to the rare heavy
to light decays $B \to K^* \ell^+ \ell^-$
and $B \to \rho \ell^+ \ell^-$
with $\ell = e, \mu$.

Data on these decays are still very sparse; there are first measurements 
of the decays $B \to \pi \ell \nu$ and  $B \to \rho \ell \nu$ from CLEO
\cite{Berkel}, from which total rates may be obtained. From this one 
may extract a value of $V_{ub}$ by employing form factor models, and the 
value given by CLEO is 
\begin{equation} 
|V_{ub}| = (2.6 \mbox{ to }4.0 \pm 0.2 ^{+0.3}_{-0.4}) \times 10^{-3}
\end{equation}
where the errors are purely experimental, while the range in the 
central value indicates the span obtained from a representative set
of models. In order to 
perform a model independent determination 
along the lines discussed above a good measurement of the lepton energy 
spectra in these decays is needed. 
\section{Conclusions}
The standard model has turned out to be suprisingly successful and 
has passed many tests, in particular the very precise tests performed 
at the LEP collider. However, these tests mainly concern the coupling 
of the gauge bosons to the fermions while the CKM sector of the 
standard model has not been tested with comparable accuracy.

To test this part of the standard model one has to investigate weak 
processes among quarks which are in general plagued with strong 
interaction uncertainties. In this respect the heavy mass limit 
has brought an enormous success; HQET opens the unique
possibility to determine some of the CKM matrix elements in a model 
independent way, thereby reducing uncertainties through models  
considerably. In particular, it allows to at least give the order
of magnitude of the uncertainties involved, since HQET relies on an 
expansion in $\alpha_s (m_Q)$ and $1/m_Q$. 

Since the discovery of the heavy quark symmetries their phenomenological
applications as well as the theoretical background have been studied 
intensively. The most prominent example is the determination of $V_{cb}$, 
which corresponds to a heavy to heavy transition. Combining the method 
as described in this mini-review fo exclusive decays with the 
$1/m_Q$ expansion for inclusive decays \cite{Bigi} one may by now 
determine $V_{cb}$ up to an uncertainty significantly less than 
ten percent. 

In heavy to light 
decays heavy quark symmetries do not work as efficiently; in this case
only the relative normalization of $B$ decays versus the corresponding 
$D$ decays may be obtained. From the experimental side there are first 
measurements of $B \to \pi \ell \nu$ and $B \to \rho \ell \nu$ from the 
CLEO collaboration and 
an extraction of the CKM matrix element $V_{ub}$ from these processes
is still to some extent model dependent. A more model independent 
extraction of this matrix element has to wait for more data, in 
particular a measurement of the lepton energy spectrum is needed to 
exploit the relative normalization from heavy quark symmetry. 

HQET does not yet have much to say about exclusive non-leptonic decays; even 
for the decays $B \to D^{(*)} D^{(*)}_s$, which involves three heavy quarks, 
heavy quark symmetries are not sufficient to yield useful relations between
the decay rates \cite{MRRnonlep}. Of course, with additional assumptions 
such as factorization one can go ahead and relate the 
non-leptonic decays to the semileptonic ones; however, this is a very strong
assumption and it is not clear in what sense factorization is an 
approximation. On the other side, the data on the non-leptonic $B$ decays 
support factorization, and first attempts to understand this from QCD and 
HQET have been untertaken \cite{DGnonlep}; however, the problem of the 
exclusive non-leptonic decays still needs clarification and hopefully the 
heavy mass expansion will also be useful here.

\section*{Figure Captions}
\begin{itemize}
\item[Fig.1] CLEO data\cite{cleovcb} and the extrapolation 
             used to obtain $V_{cb}$. The upper figure corresponds to 
             a linear extrapolation, the lower one includes also a curvature.
\item[Fig.2] ALEPH data\cite{ALEPHvcb} and the extrapolation used to 
             obtain $V_{cb}$.
\item[Fig.3] Data and the extrapolation used to obtain $V_{cb}$ from 
             $B \to D \ell \nu_\ell$. The figure is taken from 
             \cite{cleovcb1}
\end{itemize}
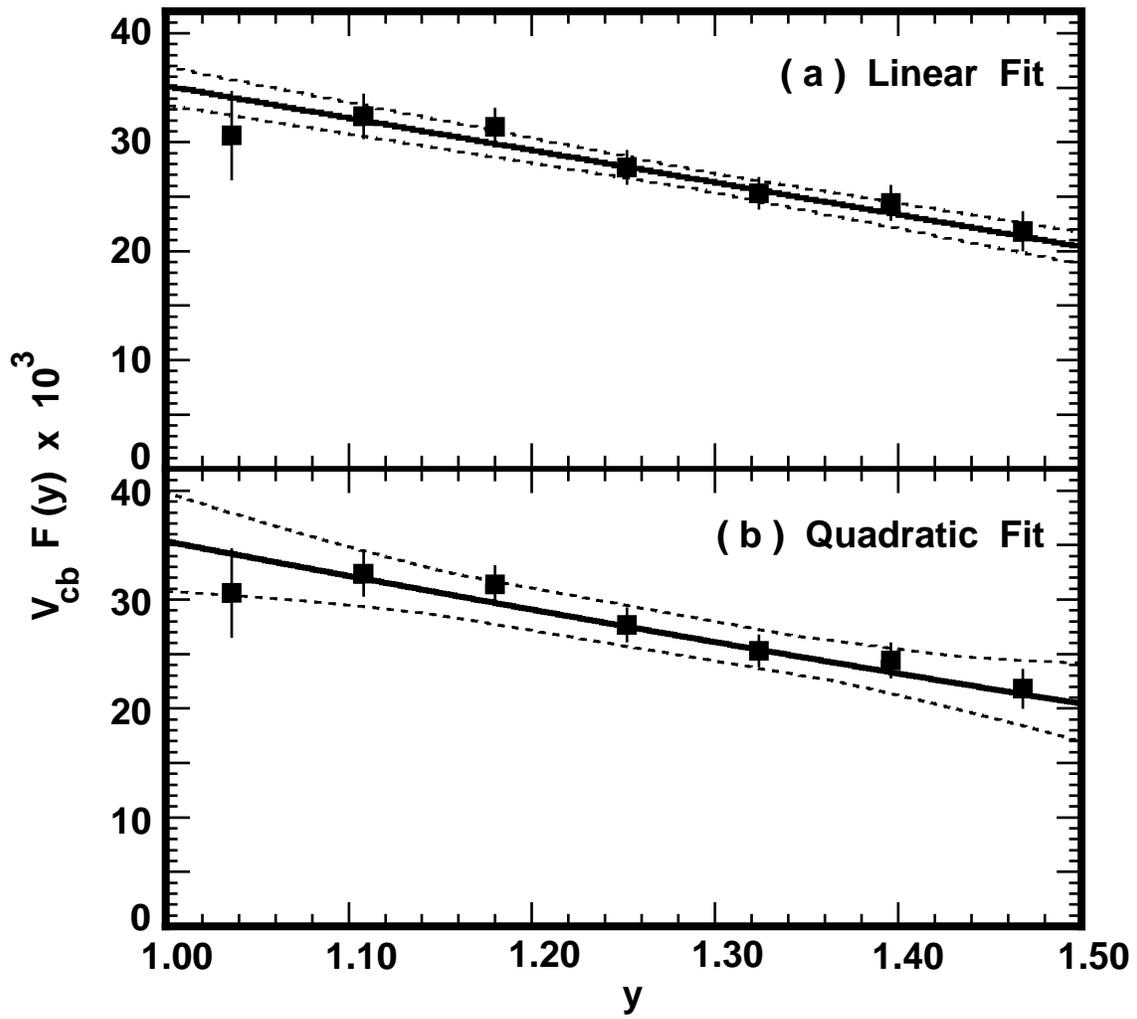
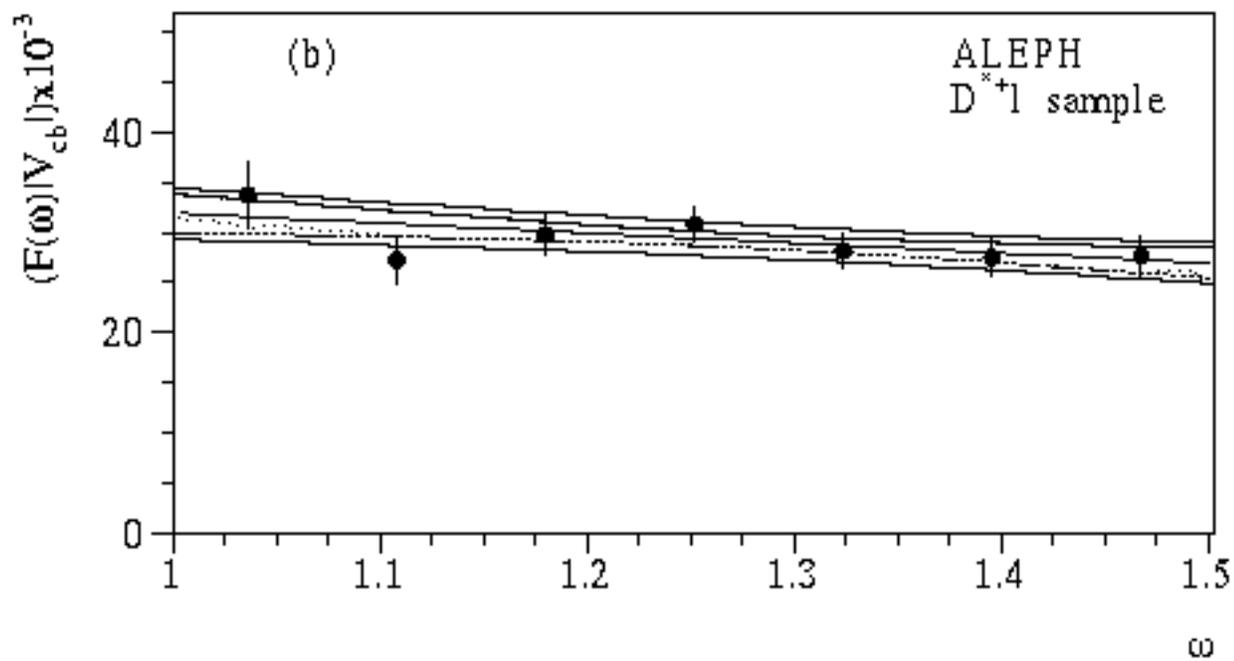
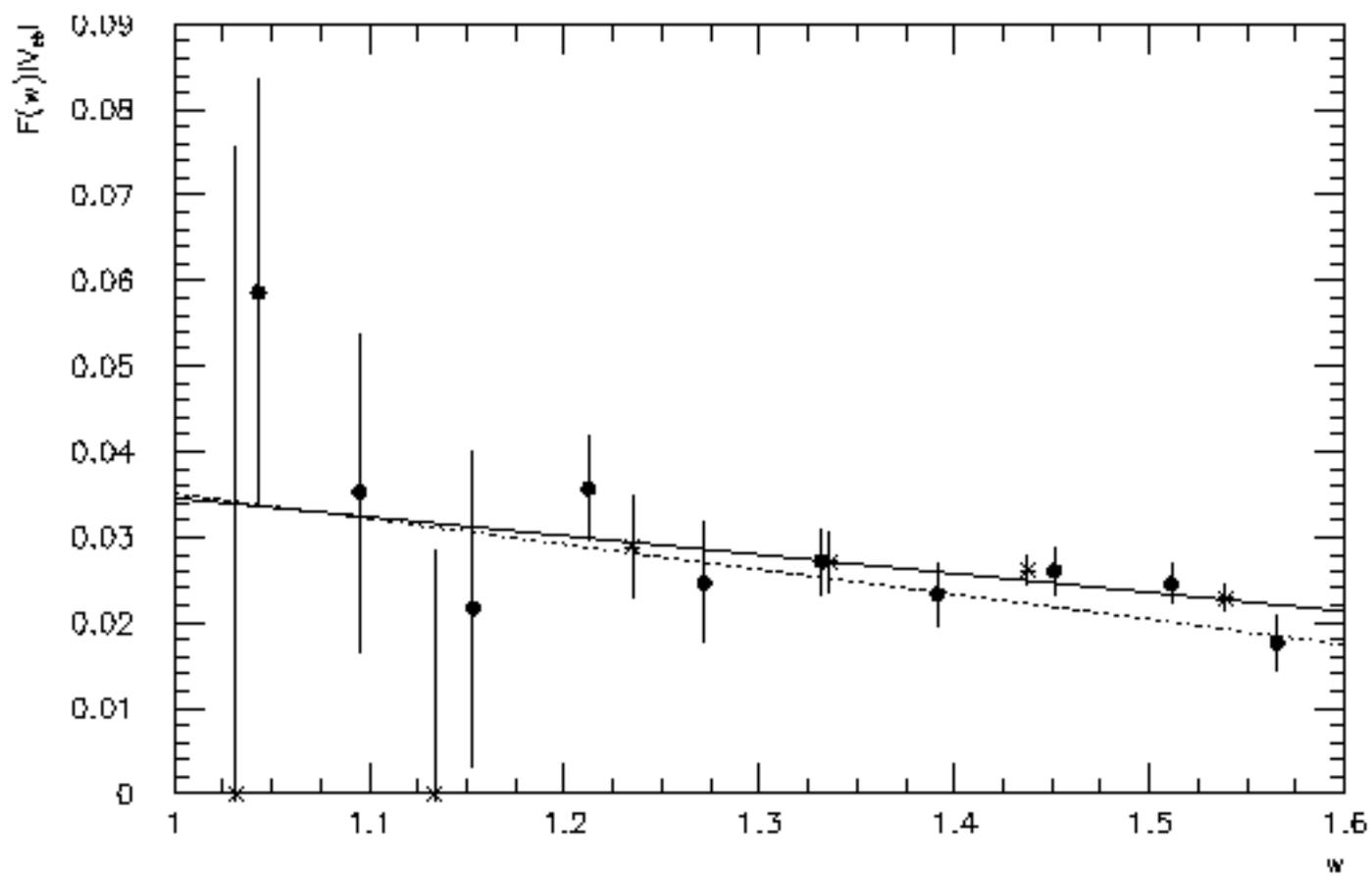
\end{document}